\documentclass[journal]{IEEEtran}

\ifCLASSINFOpdf
\else
\fi

\usepackage{setspace}
\usepackage{bbm}
\usepackage[cmex10]{amsmath}
\usepackage{amssymb}
\usepackage{cite}
\usepackage{graphicx}
\usepackage{array,color}
\usepackage{amsmath}
\allowdisplaybreaks
\usepackage{stfloats}
\usepackage{graphicx}
\usepackage{epstopdf}
\usepackage{tabularx}
\usepackage{epsfig,epsf,color,balance,cite}
\usepackage{algorithmic}
\usepackage{algorithm}
\usepackage{url}
\usepackage{bm}
\usepackage{multirow}
\usepackage{hyperref}
\usepackage{amsthm}
\usepackage{gensymb}
\usepackage[caption=false,font=footnotesize,labelfont=rm,textfont=rm]{subfig}

\ifodd 1
\usepackage{soul}
\usepackage{color}
\setstcolor{red}

\newcommand{\del}[1]{\st{#1}} %deleting the text

\newcommand{\com}[1]{\textbf{\color{red} (COMMENT: #1)}} %comment of the text
\newcommand{\response}[1]{\textbf{\color{green} (RESPONSE: #1)}} %response to comment
\else

\newcommand{\del}[1]{}

\newcommand{\com}[1]{}
\newcommand{\comg}[1]{}
\newcommand{\response}[1]{}
\fi
\title{\LARGE {Rotatable Antenna-Enabled Space-Air-Ground Integrated Networks: Opportunities and Challenges}}
\author{ 
    Yanhua Tan,
	Beixiong Zheng,~\IEEEmembership{Senior Member,~IEEE},
    Tiantian Ma,
    Qingjie Wu, 
    Lipeng Zhu,~\IEEEmembership{Senior Member,~IEEE},
    Wenyan Ma,~\IEEEmembership{Member,~IEEE},
    Cheng-Xiang Wang,~\IEEEmembership{Fellow,~IEEE}, \\
    Robert Schober,~\IEEEmembership{Fellow,~IEEE}, 
    and Rui Zhang,~\IEEEmembership{Fellow,~IEEE}
}

\begin{document}

\maketitle

\begin{abstract}
Space-air-ground integrated networks (SAGINs) are expected to support ubiquitous three-dimensional (3D) communication and sensing services in future sixth-generation (6G) wireless networks. However, heterogeneous mobility and service requirements across the space, air, and ground segments challenge fixed-boresight or fixed-sector antennas in supporting wide-area coverage, maintaining directional alignment over time-varying communication links, and flexibly adjusting observation directions for sensing tasks. To address these limitations, rotatable antenna (RA) technology has emerged as a promising solution by enabling mechanical or electronic boresight adjustment while keeping the antenna positions fixed. This article investigates the roles of RA technology in enabling communication and sensing in SAGINs. Specifically, we first explain how RA boresight control complements platform mobility and supports wide-area coverage, dynamic directional transmission, cross-segment cooperative operation, and cooperative sensing. We then discuss the main design challenges, potential approaches, and future research directions, including channel acquisition and predictive tracking, joint orientation control and access/handover coordination, cross-segment coordination and resource management, as well as deployment tradeoffs and practical implementation. Finally, an RA array prototype and two representative simulation studies are presented to illustrate the feasibility and potential performance gains of RA-enabled SAGINs.
\end{abstract}

\IEEEpeerreviewmaketitle

\section{Introduction}\label{sec:introduction}
Future sixth-generation (6G) wireless networks are expected to support not only higher data rates and lower latency, but also ubiquitous connectivity, integrated sensing, and intelligent service delivery across highly diverse environments~\cite{Wang2023On}. Achieving this vision requires future networks to move beyond purely terrestrial communication infrastructures, which are inherently constrained in providing continuous and reliable services in remote and infrastructure-limited scenarios. To address these limitations, space-air-ground integrated networks (SAGINs) have emerged as a promising architecture that tightly combines satellites, aerial platforms, and terrestrial infrastructures into a unified service framework~\cite{Xiao2024Space}. The space segment typically consists of satellites in geostationary Earth orbit (GEO), medium Earth orbit (MEO), and low Earth orbit (LEO), while the air segment includes high-altitude platforms (HAPs) and low-altitude platforms (LAPs). By exploiting the complementary capabilities of these heterogeneous segments, SAGINs can significantly enhance the coverage, flexibility, and service capability of future wireless networks.
%~\cite{Liu2018Space}

Despite these advantages, efficient SAGIN operation remains challenging due to the highly heterogeneous and dynamic nature of the network. Specifically, satellites, aerial platforms, and terrestrial infrastructures operate with different coverage ranges, mobility patterns, and propagation environments, leading to time-varying channel conditions, intermittent blockage, and frequent access and handover decisions. Meanwhile, SAGINs need to support wide-area connectivity, high-capacity access, relay and backhaul links, as well as wide-area sensing and target tracking. Although existing fixed-antenna architectures can adapt to changing service and link conditions through beamforming, power control, and resource allocation, their fixed boresight directions or predefined service areas limit spatial adaptability when service regions, link directions, and sensing tasks vary over time. Compensating for this limitation may require larger antenna arrays, denser infrastructure deployment, or more frequent beamforming updates and resource reallocation, thereby increasing hardware complexity, control overhead, and energy consumption.

%Beyond communication, satellites and aerial platforms may also need to perform wide-area sensing and track moving targets. These diverse communication and sensing requirements expose the limitations of fixed-antenna architectures. Their fixed boresight directions or predefined service areas cannot readily adapt to dynamic service demands, time-varying link conditions, and task-dependent sensing requirements. As a result, improving coverage flexibility and spatial adaptability typically relies on deploying denser infrastructures or larger antenna arrays, thereby increasing hardware complexity and energy consumption.

To improve spatial adaptability, flexible antenna architectures have recently attracted increasing attention in non-terrestrial and SAGIN-related scenarios.  In particular, fluid antenna system (FAS) has been studied for non-terrestrial networks (NTNs), where reconfigurable antenna states can help adapt to dynamic propagation and interference conditions \cite{Xu2026Advancing}. Moreover, movable antenna (MA) has been investigated for LEO satellite communications~\cite{Zhu2025Dynamic}, while six-dimensional movable antenna (6DMA), which jointly adjusts antenna positions and orientations, has been explored in aerial communication scenarios~\cite{Ren2025SixDimensional}. These studies indicate that antenna reconfigurability can improve coverage adaptability, link quality, and interference management in satellite and aerial networks. However, the existing studies mainly focus on individual links or specific network segments, while coordinated operation of flexible antennas across the space, air, and ground segments remains relatively unexplored. Moreover, MA-based position reconfiguration generally requires additional movement space, while 6DMA~\cite{Shao20256DMA} further introduces orientation adjustment and the associated control complexity. These requirements may limit compact integration on SAGIN platforms with strict payload, size, and energy constraints.

\begin{figure*}
    \centering
    \includegraphics[width=\textwidth]{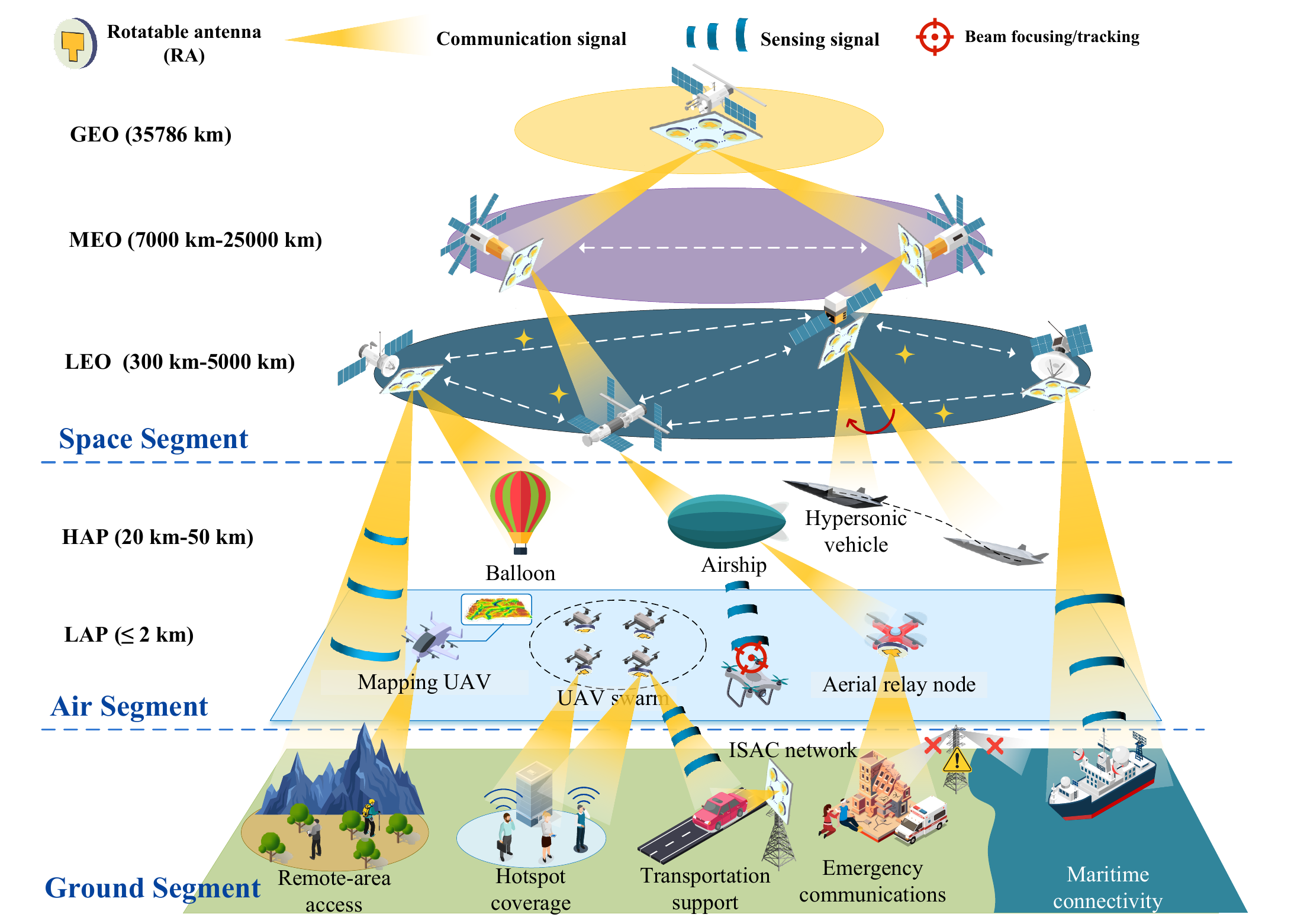}
    \caption{An architecture for RA-enabled SAGINs.}
    \label{fig:intro_scenario}
\end{figure*}

Among these emerging flexible antenna architectures, rotatable antenna (RA) provides an attractive solution for dynamic wireless networks~\cite{Zheng2026Rotatable,Wu2025Modeling,Zheng2025Rotatable,Zheng2026RotatableAntenna,Wu2026Antenna,Ma2026Rotatable}. As a rotation-centric flexible antenna architecture, RA enables spatial adaptability through mechanical or electronic boresight adjustment. 
By retaining rotational reconfigurability while keeping the antenna position fixed, RA provides a relatively compact implementation with lower hardware and control complexity. This property is attractive for SAGIN nodes subject to strict payload, size, and energy constraints. More importantly, orientation-domain reconfigurability enables the decoupling of antenna orientation from platform trajectory and attitude control, allowing satellite and aerial nodes to maintain their planned trajectories while dynamically adapting their boresight directions to changing task requirements~\cite{Wu2026Antenna,Ma2026Rotatable}. RA-equipped nodes can further exploit this capability to either scan different spatial regions for wide-area coverage and observation or track selected links and targets to maintain favorable directional gain. Such orientation adaptability enables more flexible spatial resource utilization and supports adaptive service provisioning in highly dynamic three-dimensional (3D) network environments.

As shown in Fig.~\ref{fig:intro_scenario}, RA-enabled SAGINs provide a unified 3D network framework for a variety of future 6G service scenarios, such as remote-area access, aerial coverage extension, maritime connectivity, transportation support, and emergency communications. 
%In such a network, RA-equipped satellites, aerial platforms, and ground nodes can adjust their boresight directions toward selected nodes or service regions as the network geometry evolves, thereby providing directional support for cross-segment access, relay, and backhaul links. By coordinating the boresight directions of participating nodes, RA can further facilitate the establishment and maintenance of directional links across the space, air, and ground segments. 
To clarify the roles of RA in these scenarios and the design challenges arising from practical SAGIN operation, this article provides an overview of RA-enabled SAGINs. We first discuss the communication and sensing roles of RA boresight control in mobile SAGINs. We then examine the associated design challenges, potential approaches, and future research directions. Finally, an RA array prototype and two representative simulation studies illustrate implementation feasibility and potential performance gains of RA-enabled SAGINs.
\begin{table*}[t]
\caption{Representative Tasks and Roles of RAs in SAGINs}
\label{tab:RA_SAGIN_roles}
\centering
\footnotesize
\renewcommand{\arraystretch}{1.30}
\setlength{\tabcolsep}{3.5pt}

\renewcommand{\tabularxcolumn}[1]{m{#1}}

\hyphenpenalty=10000
\exhyphenpenalty=10000
\emergencystretch=2em

\begin{tabularx}{\textwidth}{
|>{\hsize=0.55\hsize\centering\arraybackslash}X
|>{\hsize=1.15\hsize\centering\arraybackslash}X
|>{\hsize=0.90\hsize\centering\arraybackslash}X
|>{\hsize=1.40\hsize\centering\arraybackslash}X|
}
\hline

\textbf{Segment/Operation} &
\textbf{Communication Tasks} &
\textbf{Sensing Tasks} &
\textbf{Main Role of RAs} \\
\hline

Space segment &
Wide-area access and long-range links &
Regional observation and remote sensing &
Scan service regions and track long-range links \\
\hline

Air segment &
Regional coverage, relaying, and satellite backhaul &
Environmental monitoring, disaster assessment, and target tracking &
Adjust the boresight while maintaining the planned flight trajectory \\
\hline

Ground segment &
Local access, gateway connectivity, and satellite/aerial links &
Local sensing and complementary measurements &
Track moving satellites or aerial platforms \\
\hline

Cross-segment operation &
End-to-end access, relay, backhaul, and cooperative communications &
Cooperative sensing and localization/navigation support &
Coordinate boresights across participating nodes \\
\hline

\end{tabularx}
\end{table*}

\section{RA-Enabled Communication and Sensing in SAGINs}

%This section discusses how RA boresight control complements platform mobility to support communication and sensing in SAGINs.  
Since space, air, and ground nodes exhibit different mobility patterns, coverage characteristics, and service roles, the functions of RAs vary across SAGIN scenarios. Table~\ref{tab:RA_SAGIN_roles} summarizes representative communication and sensing tasks and RA roles across different segments. We next discuss RA boresight control for mobile platforms, as well as RA-enabled communication and sensing.

\subsection{RA Boresight Control for Mobile SAGIN Platforms}
In mobile SAGINs, the relative directions among satellites, aerial platforms, and ground nodes vary continuously with orbital motion, flight trajectories, and user mobility. RA provides a flexible means of adapting to these directional variations through controllable mechanical or electronic boresight adjustment while keeping the antenna positions fixed. %Unlike fixed antennas, RA allows the boresights of individual antennas or an entire antenna array to be adjusted toward desired directions in 3D space. 
This controllable boresight adjustment introduces additional spatial degrees of freedom (DoFs) that complement conventional signal-domain beamforming. This is particularly useful for mobile SAGIN platforms, whose trajectories and attitudes are often constrained by their primary missions and therefore cannot always be adjusted solely to maintain directional alignment.

%Specifically, RA adjusts the boresight direction of the antenna radiation pattern, whereas conventional beamforming adjusts the complex weights applied to the antenna signals. The two mechanisms can therefore be jointly used to align the overall antenna response with selected communication links, service regions, or sensing targets. More importantly for mobile SAGIN platforms, this additional DoF allows the antenna pointing direction to be adjusted without relying solely on platform motion or attitude control.

%This complementary relationship between platform mobility and RA boresight control is particularly valuable for satellite and aerial nodes. 

Accordingly, platform mobility and RA boresight control provide complementary forms of spatial adaptation. Platform motion determines the accessible service or observation region, while RA provides finer directional adaptation within that region. For example, an uncrewed aerial vehicle (UAV) equipped with a fixed directional antenna may need to change its body attitude or flight path to maintain alignment with a satellite, ground user, or sensing target. With RA, the UAV can follow its intended path while adjusting its antenna boresight toward the current communication link or sensing target. Similarly, a satellite can continue along its orbit while directing its onboard RA system toward a ground service region, an aerial relay, or another satellite. RA-equipped nodes can scan their boresights across different spatial directions to expand service or observation coverage, or track selected links and targets to maintain directional alignment and gain. These two operating modes, namely boresight scanning and boresight tracking, form the basis of the communication and sensing functions discussed below.

\subsection{RA-Enabled Communication in SAGINs}
From the communication perspective, RA supports wide-area coverage through boresight scanning and dynamic directional transmission through boresight tracking. RA-equipped satellites or aerial platforms can sequentially direct their antenna boresights toward different spatial regions, thereby extending the effective service region over time and mitigating persistent coverage holes. This capability is particularly useful in infrastructure-limited scenarios, where terrestrial infrastructure may be unavailable. However, boresight scanning cannot provide simultaneous high-gain coverage in all directions. A wider scanning range or a larger number of service regions generally increases the revisit interval and reduces the dwell time. Therefore, these parameters should be selected according to the user distribution, traffic demand, service priority, and delay requirements.

While boresight scanning improves broad service availability, selected high-capacity links still require sustained directional alignment as the network geometry evolves.
To maintain such alignment, RA can track the time-varying directions of inter-node and cross-segment links. Otherwise, angular mismatch may reduce the received signal power and cause rate degradation or link interruption. Satellite ephemeris can assist satellite-related boresight tracking, while planned UAV trajectories and platform positions can support aerial-link tracking. Beyond individual links, RA can also facilitate cooperative communications. For an end-to-end service involving satellite backhaul, aerial relaying, and ground access, participating nodes can coordinate their boresight directions to maintain directional support along the communication path. 
%Such tracking and coordination are particularly important for high-capacity inter-node, relay, and backhaul links, where sustained directional gain needs to be maintained while the participating nodes move.

%Beyond individual links, end-to-end SAGIN services often involve satellite backhaul, aerial relaying, and ground access. The participating nodes should therefore coordinate their antenna boresights according to the selected access, relay, and backhaul links. For example, an aerial relay may use different RAs or subarrays to support the satellite backhaul link and ground users, or serve these directions in different time intervals. Improving one link may provide limited end-to-end benefit if another link along the same service path lacks sufficient directional gain. Therefore, RA boresight control should support the complete communication path rather than optimize each link independently.

\subsection{RA-Enabled Sensing in SAGINs}

In addition to communication, RA-equipped SAGIN nodes can support sensing through boresight scanning and tracking. Specifically, for wide-area sensing, RA-equipped satellites and aerial platforms can scan their antenna boresights across different spatial regions to support target discovery, environmental monitoring, disaster assessment, and remote sensing. In this process, platform mobility determines the broad region accessible for observation, while RA boresight control selects the instantaneous observation direction within that region. Once a target or region of interest is identified, the node can redirect its antenna boresight toward the estimated target direction and update it as the platform or target moves. 
%Therefore, boresight scanning expands observation coverage, whereas boresight tracking maintains directional support for selected targets or regions of interest.
The resulting directional measurements can also support positioning, target localization, and navigation assistance when satellites, aerial platforms, or ground nodes provide reference signals. 
For these tasks, RA facilitates signal acquisition from selected directions, while the achievable accuracy also depends on reference-node geometry, synchronization, and signal processing.

Beyond single-node observation, RA can further support cooperative sensing across the space, air, and ground segments. In particular, satellites can provide broad regional observation and remote sensing information, aerial platforms can offer flexible and shorter-range sensing with rapidly adjustable observation directions, and ground nodes can provide complementary local measurements. 
By coordinating their boresight directions, these nodes can observe the same target or region from different altitudes and directions, thereby providing complementary observations and improving robustness to blockage.
%By coordinating their boresight directions, these nodes can observe the same target or region from different altitudes and directions, thereby reducing the impact of blockage and maintaining observation continuity when one node no longer has a favorable observation direction. 
Such cooperation relies on communication links for exchanging sensing data, task information, and control messages. Meanwhile, sensing results, such as target locations and blockage information, can assist communication coverage adaptation, boresight tracking, and relay selection. Conversely, user locations, platform trajectories, and link information can help narrow the sensing search region and guide boresight adjustment.

\section{Main Design Challenges, Potential Approaches, and Future Directions}
RA-enabled SAGINs face several design challenges in practical systems. This section discusses representative approaches and future research directions for channel acquisition and predictive tracking, joint orientation control and access/handover coordination, cross-segment resource management, and deployment tradeoffs and practical implementation.

\subsection{Channel Acquisition and Predictive Tracking}
Accurate channel and directional information is essential for effective RA boresight control in SAGINs. A key challenge is that the channel may continue to evolve while the RA boresight is being adjusted, especially for high-mobility satellite and aerial links. When the RA reconfiguration time is sufficiently shorter than the channel coherence time, the channel can be approximately treated as quasi-static during each boresight update. However, in highly dynamic SAGIN environments, orbital motion, flight trajectories, user mobility, and blockage variations may cause the link geometry and channel state to change before the boresight adjustment is completed. Consequently, previously acquired channel information may become outdated, leading to channel aging, boresight mismatch, and degraded link performance. Nevertheless, dominant geometric information, such as the main propagation direction, often evolves more smoothly and predictably than the instantaneous small-scale fading, particularly for line-of-sight (LoS)-dominant satellite and aerial links~\cite{Wang2025Modeling}. Therefore, relatively stable geometric information can still be exploited as prior information for link-direction prediction, thereby reducing repeated channel acquisition.

%Therefore, RA boresight control can exploit such relatively stable geometric information instead of relying on repeated full-channel estimation for every update.

To reduce channel acquisition overhead and cope with channel aging, predictive tracking can exploit prior information available in SAGINs. Specifically, satellite ephemeris, planned aerial trajectories, node locations, and previously acquired channel or directional observations can be used to predict link directions and guide subsequent boresight updates. For links with relatively predictable geometric evolution, trajectory- and location-assisted prediction can narrow the directional search range. For more general dynamic links, local angular observations can refine the predicted directions when trajectory deviations, blockage variations, or other unexpected changes reduce the prediction accuracy. In addition, RA boresight adjustment can provide directional observations from different orientations, which may help resolve directional uncertainty without repeatedly acquiring the complete channel. Moreover, the prediction horizon should account for the RA reconfiguration delay to ensure that the predicted link direction remains valid when the boresight update is completed. A remaining challenge is determining when the available prediction is sufficiently reliable and when additional channel or directional observations should be triggered. Future research may investigate adaptive acquisition and tracking schemes that balance prediction accuracy, observation overhead, and RA reconfiguration latency.

\subsection{Joint Orientation Control and Access/Handover Coordination}
In RA-enabled SAGINs, antenna orientation cannot be designed independently of access selection and handover. Besides conventional factors such as link quality and segment availability, service performance depends on whether the antenna boresight is properly aligned with the corresponding link direction. Consequently, access, handover, and antenna orientation are tightly coupled. As the serving node or network segment changes, coordinated boresight adjustment may therefore be required along the affected access, relay, and backhaul links. For highly directional RA-enabled links, the boresight should be aligned with the direction of the new serving node before or during the handover so that sufficient directional gain is available when the new link becomes active. Otherwise, switching to a nominally available link before directional alignment is established may cause transient link degradation or even service interruption. 

Antenna orientation also influences access and relay selection because the effective service region of an RA-equipped node varies with its current boresight. 
%Therefore, antenna orientation affects not only the execution of a handover but also the selection of the serving node and service path. 
To address this coupling, RA-enabled SAGINs require joint design of antenna orientation, access selection, and handover coordination. A practical approach is to consider these decisions over different timescales. At a slower timescale, the network can exploit large-scale  information, such as user distribution, service demand, orbital evolution, and platform trajectories, to determine candidate serving nodes, relay paths, and coarse boresight configurations. At a faster timescale, local boresight adjustment can maintain directional alignment with the selected link and support timely handover execution as the service geometry changes. In this way, RA-enabled SAGINs can better maintain directional support along the selected service path while limiting handover overhead and reducing the risk of service interruption. An important future direction is to further exploit multiple RAs or subarrays to support seamless handover. For example, part of the antenna resources may maintain the current link while the remaining RAs establish directional alignment with a candidate serving node before switching. How to coordinate the directional support for the current and candidate links under limited antenna resources and rapidly changing link geometry remains an open problem.
\subsection{Cross-Segment Coordination and Resource Management}
Building on joint orientation control and access/handover coordination, cross-segment resource management further determines how the limited network resources should be allocated across the selected access, relay, and backhaul links. Conventional SAGINs already need to coordinate spectrum, transmit power, sensing resources, and backhaul capacity. However, when RAs are introduced, resource allocation also depends on antenna orientation, since the effectiveness of the assigned spectrum and power depends on whether the corresponding links can obtain sufficient directional gain. Multiple communication and sensing tasks may also compete for a limited number of RAs, subarrays, or orientation states. Meanwhile, overlapping links may cause mutual interference. Since the interference level also depends on the relative boresight directions of the involved nodes, antenna orientation introduces an additional spatial resource that affects both the desired-link enhancement and interference coupling. 
%Therefore, cross-segment resource management needs to jointly account for conventional resource bottlenecks, limited directional support, and interference among simultaneously active links.

To address these challenges, limited directional support can be alleviated through boresight adjustment or subarray reassignment, whereas bandwidth or backhaul limitations may require traffic redirection through another segment or relay.
%resource allocation should first identify the main bottleneck along the selected access, relay, and backhaul links. The bottleneck may arise from insufficient bandwidth, constrained backhaul capacity, limited transmit power, heavy traffic load, or inadequate directional gain.
%The network can identify whether the main bottleneck comes from insufficient bandwidth, constrained backhaul capacity, limited transmit power, heavy traffic load, or inadequate directional gain. 
%When directional gain is the main limitation, boresight adjustment, subarray reassignment, or time-scheduled orientation control can be considered before allocating more spectrum or power. When bandwidth or backhaul capacity is the bottleneck, traffic can be shifted to another segment, routed through an alternative relay, or scheduled over different transmission periods, provided that the new links can still obtain sufficient directional support. 
For interference circumvention, the boresight directions of RA-equipped nodes can be adjusted to increase the antenna gain toward desired links while steering lower-gain regions toward dominant interfering directions, thereby reducing mutual interference. When the desired and interfering directions are sufficiently separated, such directional adjustment can further facilitate spectrum reuse; otherwise, power control, orthogonal spectrum allocation, or time-domain separation may still be required. 
%Moreover, sensing observations and network feedback can help identify service hotspots, blocked regions, coverage holes, and interference-prone directions, thereby guiding load balancing, resource allocation, and boresight assignment across different segments. 
A remaining challenge is that cross-segment coordination relies on timely information about traffic load, link quality, interference, and RA boresight directions. Acquiring and exchanging such information can introduce considerable signaling overhead, while the information may quickly become outdated in highly dynamic SAGINs. Future research may explore distributed or hierarchical coordination methods that rely mainly on local information and limited cross-segment information exchange to coordinate network resources and RA boresight control.

\subsection{Deployment Tradeoffs and Practical Implementation}
Although RA provides a relatively compact means of introducing spatial adaptability into SAGINs, its practical deployment still needs to balance implementation feasibility with network-level performance gains. In particular, satellites require high reliability and stable calibration under strict payload and power budgets, while aerial platforms are more sensitive to size, weight, energy consumption, and platform vibration. Ground nodes usually allow easier installation and maintenance but still require cost-efficient and scalable deployment. Accordingly, the boresight adjustment mechanism, deployment scale, and update frequency should be selected according to platform capability, service role, and link dynamics.

At the hardware level, a key tradeoff lies in the choice of boresight adjustment mechanism. In mechanically driven RA designs, physical rotation directly adjusts the antenna boresight in 3D space. This approach generally provides a wide steering range, making it suitable for coarse pointing and slow-timescale boresight tracking over links with predictable geometric evolution~\cite{Baek2003AVband}. However, frequent mechanical adjustment may introduce non-negligible actuation latency, mechanical wear, and maintenance burden. By contrast, electronically driven boresight adjustment keeps the antenna physically fixed and adjusts the radiation pattern through feed selection, impedance tuning, or parasitic-element control~\cite{Zhang2022Highly}. This enables faster directional adjustment and easier integration with compact platforms, but the available steering directions are often discrete and predefined, and the synthesized patterns may require careful calibration across different electronic states. As a further extension, hybrid RA architectures can combine wide-angle mechanical pointing with fast electronic refinement, but they also increase hardware complexity and calibration overhead. %Therefore, the boresight adjustment mechanism should be matched to the segment dynamics and platform constraints, rather than selected as a uniform hardware solution.

Beyond the hardware mechanism, deployment scale and update frequency also affect practical operation. RAs may be installed only at critical nodes, such as satellites, aerial relays, or gateways, to support key links with lower cost and control overhead. In contrast, wider or distributed deployment can improve coverage flexibility and network robustness but increases calibration and cross-segment coordination overhead. Similarly, frequent boresight adjustment may enhance directional support, but excessive updates can introduce additional signaling burden and reconfiguration latency. Therefore, a feasible strategy is to maintain coarse directional support through relatively stable orientations and trigger finer adjustment only when relative geometry, traffic demand, blockage, or interference conditions change significantly. However, practical RA performance is also affected by hardware constraints, including reconfiguration latency, orientation accuracy, calibration errors, and energy consumption. Future research should investigate how these constraints affect communication and sensing performance and how they can be incorporated into RA control and deployment design. %Larger-scale prototypes and field trials are also needed to evaluate the resulting network-level performance gains under realistic operating conditions.

\begin{figure*}[t]
	\vspace{-0.1cm}
	\center
	\includegraphics[width=0.95\linewidth]{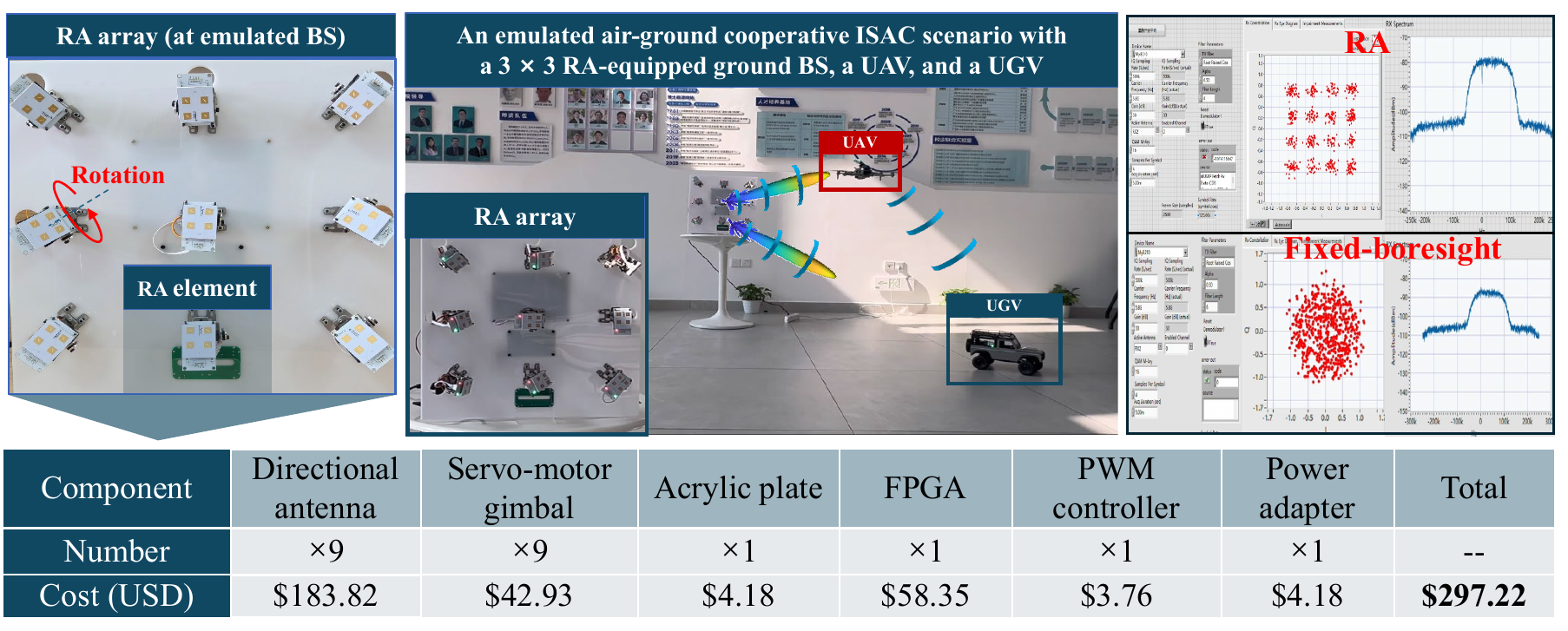}
	\vspace{-0.1cm}
	\caption{RA array prototype for air-ground cooperative ISAC.}\vspace{-0.1cm}
	\label{RA_prototype_array}  %label for entire figure
	\vspace{-0.0cm}
\end{figure*}

\section{Case Studies}
This section presents an RA array prototype and two representative simulation studies for air-ground integrated sensing and communication (ISAC) and dynamic ground-satellite communication.

\begin{figure}[h!]
    \centering   
     \includegraphics[width=0.9\columnwidth]{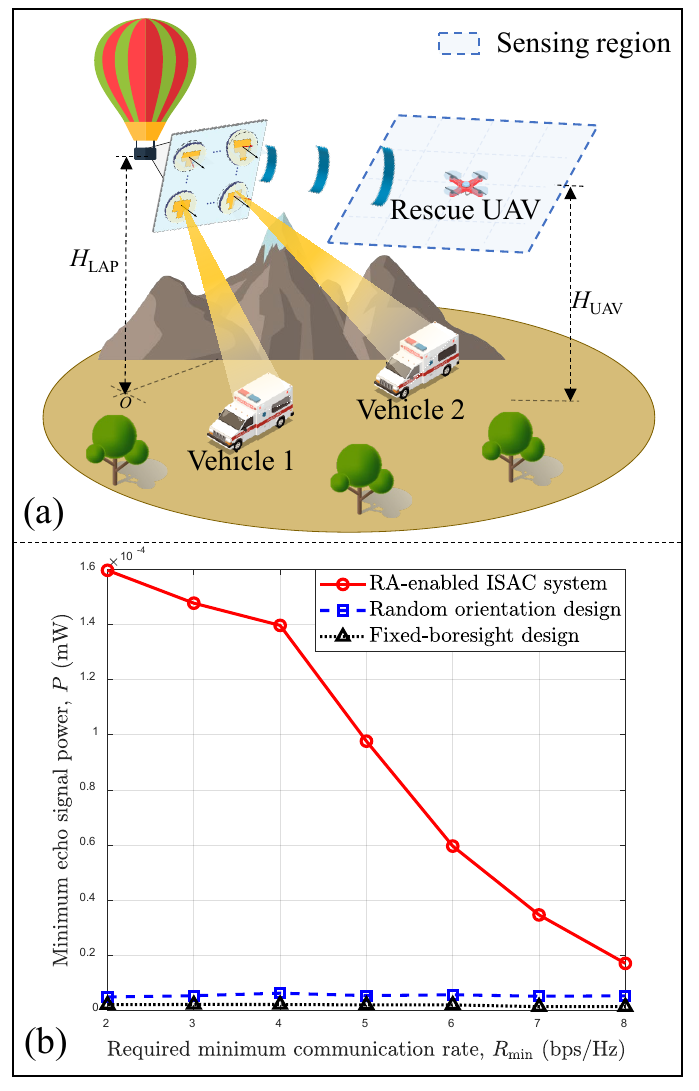}
    \caption{RA-enabled air-ground ISAC for emergency response.}
    \label{fig:air_ground_isac}
\end{figure}
%(a) An RA-equipped LAP communicates with two emergency-response vehicles while sensing an aerial region in which a rescue UAV may be located. (b) Minimum echo signal power over the aerial sensing region versus the required minimum communication rate.
\subsection{RA Array Prototype for Air-Ground Cooperative ISAC}
To validate the practical feasibility of RA arrays in air-ground cooperative ISAC, Fig.~\ref{RA_prototype_array} illustrates an experimental setup based on a $3\times3$ RA array. A short demonstration video is available online.\footnote{Available at: \href{https://www.youtube.com/watch?v=3DQIkI4ceFE}{https://www.youtube.com/watch?v=3DQIkI4ceFE}} In this experiment, the RA array operates as a ground base station (BS), while terminal emulators are deployed on a UAV and an uncrewed ground vehicle (UGV). Based on their locations, the ground BS can group and configure the RAs as spatial resources, with different groups adjusting their boresight directions toward the UAV and UGV according to the communication and sensing tasks. Furthermore, the UAV can also serve as an auxiliary aerial sensing node to assist the ground BS in acquiring the UGV position when ground-side sensing is limited by blockage. This cooperative operation illustrates how RA grouping, boresight adjustment, and UAV-assisted sensing can be integrated within an air-ground ISAC prototype to support communication and sensing tasks involving aerial and ground nodes.

Notably, the prototype also indicates the potential for low-cost implementation of RA arrays. As summarized in Fig.~\ref{RA_prototype_array}, the total hardware cost is only 297.22 USD, including the directional antennas, servo gimbals, a field-programmable gate array (FPGA), a pulse-width modulation (PWM) controller, and a power adapter. This highly competitive cost structure not only highlights the practicality and efficiency of the RA architecture but also underscores that it is a cost-effective and scalable solution for large-scale deployment in future SAGINs.

\subsection{Simulation Results for RA-Enabled Air-Ground ISAC}

To further illustrate the benefits of RAs for air-ground ISAC, we consider an emergency-response scenario with limited terrestrial infrastructure. As shown in Fig.~\ref{fig:air_ground_isac}(a), an RA-equipped LAP communicates with two emergency-response vehicles while sensing an extended aerial region in which a rescue UAV may be located. The LAP is deployed at a height of $50$ m and equipped with a $3\times3$ RA array with half-wavelength antenna spacing. 
In a Cartesian coordinate system with the ground plane at $z=0$, the LAP is located at $(0,0,50)^T$ m, while the two vehicles are located at $(-20,35,0)^T$ m and $(20,40,0)^T$ m, respectively.
The aerial sensing region is an $8\times8$ m$^2$ square centered at $(0,35,30)^T$ m and is uniformly discretized into $9\times9$ sampling points.
The system operates at $2.4$~GHz, while the maximum RA rotation angle and antenna directivity factor are set to $\theta_{\max}=\pi/6$ and $\rho=4$, respectively. 
%The directivity factor $\rho$ reflects the antenna directivity and mainlobe beamwidth, with a larger $\rho$ corresponding to a narrower mainlobe and a more concentrated directional gain around the antenna boresight. 
The maximum transmit powers for the communication and probing signals are set to $20$ dBm and $40$ dBm, respectively, and the noise power is $-80$ dBm. 
The RA boresight directions and transmit ISAC signal design are jointly optimized to maximize the minimum echo signal power among all sampled UAV locations in the aerial sensing region, which characterizes the worst-case sensing performance over the region, while satisfying a minimum communication-rate requirement for each vehicle.

\begin{figure}[t]
    \centering
    \includegraphics[width=0.948\columnwidth]{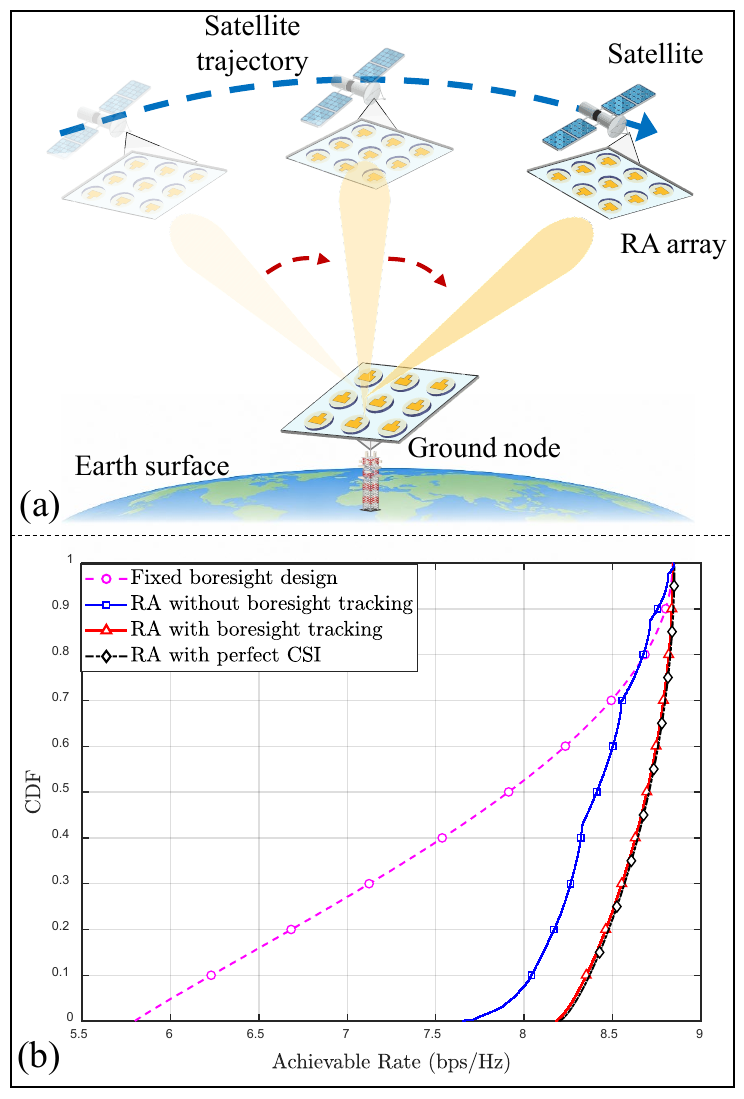}
    \caption{RA-enabled ground-satellite communication over a representative LEO satellite pass.}
    \vspace{-4mm}
    \label{fig:ground_satellite_cdf}
\end{figure}
%(a) Time-varying ground-satellite geometry and RA boresight tracking. (b) CDF of the achievable rate under different boresight update strategies.
Fig.~\ref{fig:air_ground_isac}(b) shows the minimum echo signal power achieved by the considered systems versus the required minimum communication rate $R_{\min}$. As $R_{\min}$ increases, the minimum echo signal power decreases because the more stringent communication requirements require the transmit signal design and RA boresight directions to provide stronger support to the two vehicle links, leaving less design flexibility for illuminating the aerial sensing region. As can be observed, the RA-enabled ISAC system consistently outperforms the random orientation and fixed-boresight designs by achieving higher minimum echo signal power under the same communication-rate requirement. This is because the RAs can adaptively reconfigure the array directional gain pattern by adjusting their orientations/boresights according to the locations of the two vehicles and the possible locations of the rescue UAV within the aerial sensing region. Such directional adaptation enables the system to better support effective illumination of the aerial sensing region while maintaining the two vehicle communication links.

\subsection{Simulation Results for RA-Enabled Ground-Satellite Communication}

To illustrate the role of RAs in dynamic ground-satellite communication, we consider a ground node communicating with an LEO satellite over a representative satellite pass, as shown in Fig.~\ref{fig:ground_satellite_cdf}(a). The satellite moves along a circular orbit at an altitude of 600 km, corresponding to a radius $R_{\mathrm{O}}=6.97\times10^{6}$~m and an orbital period of approximately 96 minutes, while both the ground node and the satellite employ $3\times3$ RA arrays. The carrier frequency is set to $2$~GHz, and the data-transmission frame duration is $T_{\mathrm{D}}=15$~s. The achievable rate is uniformly sampled over the considered interval to obtain its cumulative distribution function (CDF), which represents the fraction of time for which the achievable rate does not exceed a given value.  We compare four boresight update strategies. These include: 1) \emph{RA with boresight tracking}, where the antenna boresight directions are updated according to the estimated satellite direction; 2) \emph{RA without boresight tracking}, where the antenna boresight directions are updated only during training intervals and then kept fixed during data transmission; 3) \emph{fixed-boresight design}, where the antenna boresight directions are preset and kept fixed throughout the considered interval; and 4) \emph{RA with perfect channel state information (CSI)}, which serves as an upper bound.

Fig.~\ref{fig:ground_satellite_cdf}(b) shows that the CDF curve of RA with boresight tracking achieves a more favorable rate distribution than the no-tracking and fixed-boresight designs while remaining close to the perfect-CSI upper bound. By exploiting the predictable evolution of the ground-satellite link direction, the RA boresights can be updated to follow the expected satellite direction during data transmission. In contrast, RA without boresight tracking allows directional mismatch to accumulate during each data-transmission interval, while the fixed-boresight design provides favorable alignment only for a limited portion of the pass. These results demonstrate that prediction-assisted RA boresight tracking can effectively exploit the predictable link evolution in satellite communication to reduce directional mismatch and improve the overall achievable-rate distribution.
%By updating the antenna boresights according to the time-varying ground-satellite link direction, this scheme maintains favorable directional alignment over a larger portion of the satellite pass. In contrast, RA without boresight tracking experiences a higher probability of low achievable rates because the boresight mismatch gradually increases during each data-transmission interval. The fixed-boresight system achieves the lowest performance since its preset high-gain directions can only match the satellite direction over a limited portion of the pass. These results demonstrate that timely RA boresight tracking improves not only the instantaneous link performance but also the overall achievable-rate distribution during dynamic ground-satellite communication.

\section{conclusion}
In this article, we investigated how RA can enhance communication and sensing in SAGINs by providing controllable boresight adjustment that complements platform mobility. We also discussed the key design challenges, potential approaches, and future research directions related to channel acquisition and predictive tracking, joint orientation control, access selection and handover coordination, cross-segment coordination and resource management, and practical implementation. An RA array prototype and two representative simulation studies illustrated the feasibility of RA implementation and its potential benefits for air-ground ISAC and dynamic ground-satellite communication. These discussions highlight the potential of RA as a compact, cost-effective, and scalable enabler for future agile, resilient, and intelligent SAGINs.

\ifCLASSOPTIONcaptionsoff
  \newpage
\fi

\bibliographystyle{IEEEtran}
\bibliography{RA_SAGIN}

\section*{Biographies}
\noindent{{\bf Yanhua Tan}
	(bctanyanhua06@mail.scut.edu.cn) is with the School of Microelectronics, South China University of Technology, Guangzhou, China.
}
\\

\noindent{{\bf Beixiong Zheng}
	 (bxzheng@scut.edu.cn) is a Professor with the School of Microelectronics, South China University of Technology, Guangzhou, China.
 }
\\

\noindent{{\bf Tiantian Ma}
	(mitiantianma@mail.scut.edu.cn) is with the School of Microelectronics, South China University of Technology, Guangzhou, China.
}
\\

\noindent{{\bf Qingjie Wu}
	(miqjwu@mail.scut.edu.cn) is with the School of Microelectronics, South China University of Technology, Guangzhou, China.
}
\\

\noindent{{\bf Lipeng Zhu}
	(zhulp@bit.edu.cn)  is a Professor with the School of Interdisciplinary Science, Beijing Institute of Technology, Beijing, China, and also with the State Key Laboratory of CNS/ATM, Beijing, China.
}
\\

\noindent{{\bf Wenyan Ma}
	(wenyan@u.nus.edu) is a Research Fellow with the Department of Electrical and Computer Engineering, National University of Singapore, Singapore.
}
\\

\noindent{{\bf Cheng-Xiang Wang}
	(chxwang@seu.edu.cn) is a Professor with the National Mobile Communications Research Laboratory, Southeast University, Nanjing, China, and also with Purple Mountain Laboratories, Nanjing, China. 
}
\\

\noindent{{\bf Robert Schober}
	(robert.schober@fau.de)  is an Alexander von Humboldt Professor and the Chair for Digital Communications, Friedrich-Alexander University Erlangen-N$\ddot{\mathrm{u}}$rnberg (FAU), Erlangen, Germany. 
}
\\

\noindent{{\bf Rui Zhang}
	(elezhang@nus.edu.sg) is a Provost’s Chair Professor with the Department of Electrical and Computer Engineering, National University of Singapore, Singapore.
}
\\

\end{document}